\documentclass[a4paper,superscriptaddress,nofootinbib,10pt]{revtex4}
\usepackage{amsfonts}
\usepackage{amssymb}
\usepackage{amsmath}
\usepackage{stmaryrd}
\usepackage{color}
\usepackage{theorem}
\newtheorem{theorem}{Theorem}[section]
\newtheorem{proposition}{Proposition}[section]
\def\A{\mathcal{A}}
\def\H{\mathcal{H}}

\begin{document}

\title{Twisted bialgebroids versus bialgebroids from a Drinfeld twist
}
\author{Andrzej Borowiec}
\email{andrzej.borowiec@ift.uni.wroc.pl}
\affiliation{Institute of Theoretical Physics, University of Wroclaw, pl. M. Borna 9,
50-204 Wroclaw, Poland.}
\author{Anna Pacho{\l }}
\email{a.pachol@qmul.ac.uk}
\affiliation{Queen Mary, University of London, School of Mathematics, Mile End Rd., London E1 4NS, UK.}

\begin{abstract}
Bialgebroids (resp. Hopf algebroids) are bialgebras (Hopf algebras) over noncommutative rings. Drinfeld twist techniques are particularly useful in the
(deformation) quantization of Lie algebras as well as underlying module
algebras (=quantum spaces). Smash product construction combines
these two into the new algebra which, in fact, does not depend on the twist.
However, we can turn it into bialgebroid in the twist dependent way.
Alternatively, one can use Drinfeld twist techniques in a category of
bialgebroids. We show that both techniques indicated in the title: twisting
of a bialgebroid or constructing a bialgebroid from the twisted bialgebra give
rise to the same result in the case of normalized cocycle twist. This can be
useful for better description of a quantum deformed phase space.
We argue that within this bialgebroid framework one can justify
the use of deformed coordinates (i.e. spacetime noncommutativity)
which are frequently postulated in order to explain quantum gravity effects.
\end{abstract}

\maketitle

\section{Introduction}

Quantum groups and Hopf algebras over the years have proved their important
role in approach to Quantum Gravity. They are considered as one of the tools
of the Noncommutative Geometry which introduces more general idea of the geometry
and allows for a natural quantization of manifolds (e.g. spacetime). In some
recent works \cite{Zagreb}, \cite{LSW} the idea of generalization of Hopf
algebras into Hopf algebroids \cite{Lu,Xu} in the quantum spacetimes context
have been approached.
It is rather well known that the unification of spacetime coordinates with
a Lie algebra of symmetries cannot be done within the category of Lie
algebras. The corresponding construction is called smash (or more generally cross) product
which requires introducing the Hopf algebra framework. The algebra of
spacetime coordinates is then Hopf module algebra. A special example of this
construction is provided by the so-called phase space algebra (as the
Heisenberg algebra in Quantum Mechanics). The phase space algebra cannot be
equipped with the Hopf algebra structure (at most it can be made into
unital-non-counital bialgebra). However such smash product can be
generalized into the bialgebroid. This approach leads to more sophisticated  algebraic structures as Hopf algebroids \cite{Lu},\cite{Xu},\cite{Sweedler},%
\cite{Takeuchi},\cite{BS}. Recently the quantum
(deformed) phase spaces with the noncommutative coordinates equipped with
the bialgebroid (Hopf algebroid) structures gained some attention in the
mathematical physics literature.

Deformed quantum phase spaces have been considered shortly after the
noncommutative spacetimes were introduced, especially in the context of the
$\kappa$-deformation \cite{Luk1}. A natural extension of the $\kappa$-Poincare
quantum group \cite{Luk1,SM} by the $\kappa$-Minkowski commutation relations
contains deformation of the Heisenberg subalgebra (phase space). There have been
many constructions of such deformed phase space, e.g. within Heisenberg
double construction \cite{defPS} or smash product construction \cite{Mas,sigma6}.
Deformed quantum phase spaces are constantly studied with a
number of interesting papers appearing recently, like e.g.  \cite{Arzano}.

Focusing on the special case of deformed
quantum phase spaces with the noncommutative coordinates satisfying the $%
\kappa $-deformed Minkowski algebra \cite{Zak,SM},
the Hopf algebroid structure was firstly investigated in \cite{Zagreb}.
Later on the more detailed study on the covariance of such space under the
action of the $\kappa $-deformed Poincare symmetry within the Hopf algebroid
framework was proposed with the Heisenberg double construction naturally
providing the bialgebroid structure \cite{LSW}. Still the physical meaning
of Hopf algebroids stays not entirely clear in this context and requires more studies in
this direction. We believe  that within the bialgebroid framework one can better
justify the deformed Casimir relations which are frequently used in order to explain some quantum gravity effects coming from the spacetime noncommutativity.

Hopf algebroids are Hopf algebras over unital noncommutative rings. One of the
oldest definitions of bialgebroids go back to Sweedler \cite{Sweedler} and
Takeuchi \cite{Takeuchi}.  Schauenberg \cite{Sch} also contributed to the topic with his paper on bialgebras over the noncommutative rings. However the concept of Hopf algebroids (bialgebroids with an antipode) was introduced by Lu \cite{Lu} in 1996. The concept of twisted bialgebroids was firstly considered by Xu in 2000 \cite{Xu}. Lu's definition came as a result of
work on grupoids in Poisson geometry whereas the one by Xu came from quantum
universal enveloping algebroids (quantum groupoids) point of view. It was
later shown \cite{BM} that these two definitions are equivalent. In 2004,
G. B\"{o}hm and K. Szlachanyi \cite{BS} considered pairs of bialgebroids on which the antipode map was defined.

We will be interested in the Drinfeld twist techniques which are
particularly useful in the (deformation) quantization of (complex or real) Lie algebras
as well as underlying module algebras (=quantum spaces). The smash product
construction combines these two into a new algebra which, in fact, remains isomorphic to itself under twisting. However, if this algebra is turned into a bi- (or Hopf) algebroid, the latter is no longer isomorphic under twisting.

Our aim in this note is to show the equivalence between the bialgebroid
obtained as a result of the smash product of a twisted triangular bialgebra with
the twisted braided commutative module algebra \cite{BM} and the one obtained
as a result of twisting of the smash product in the category of bialgebroids \cite{Xu}.

\section{Preliminaries and Notations}

In this note we shall work in a category of $K-$modules, where $K$ is a base
commutative ring with unit $1\equiv 1_K$. Therefore all objects are by
default $K-$modules, all maps are $K-$linear maps. The tensor $\otimes$
product if not indicated otherwise is over the ring $K$. Particularly
interesting cases are when $K=\mathbb{K}$ is a field (of characteristic $0$)
or $K=\mathbb{K}[[h]]$ is a (topological)   ring of formal power series in the (formal)
variable $h$.
All rings (algebras) are assumed to be unital. All modules and module maps
are assumed to respect the unit. Below, for the notational convenience, we shall
briefly introduce the main notions involved in our presentation. For deeper
study we refer the reader to the literature.

\subsection{Smash Product Construction \protect\cite{Klimyk, Kassel,Majid2}%
}

Let $\mathcal{H}=(H,\Delta ,\epsilon ,1_{H})$ be a bialgebra and $\mathcal{A}%
=(A,\star ,1_{A})$ be a left $H$-module algebra with the action $\rhd
:H\otimes A\rightarrow A$ such that: $1_{H}\rhd a=a$, $M\rhd 1_{A}=\epsilon
(M)1_{A}$. Smash product algebra $A\rtimes H$ is an algebra determined on
the vector space $A\otimes H$ by the multiplication $(a\otimes L)(b\otimes
J)=a(L_{(1)}\rhd b)\otimes L_{(2)}J$, where $a,b\in A$;  $L,J\in H$ and $%
\Delta (L)=L_{(1)}\otimes L_{(2)}$ in Sweedler shortcut notation. Obviously,
the algebra $A\rtimes H$ contains algebras $A\ni a\mapsto a\otimes 1\in
A\rtimes H$ and $H\ni L\mapsto 1\otimes L\in A\rtimes H$ as subalgebras.
Later on we shall denote by $a\rtimes L$ elements from $A\rtimes H$ of the form $a\otimes L$.
Therefore the previous formula can be rewritten as
\begin{equation}
(a\rtimes L)(b\rtimes J)=a(L_{(1)}\rhd b)\rtimes L_{(2)}J.  \label{cr}
\end{equation}%
A special case of this construction provides the algebra of canonical
commutation relations between commuting coordinates and momenta generators
(see e.g. \cite{BP12} and references therein), which are fundamental from
the point of view of Quantum Mechanics (quantum phase space).
In the physically motivated examples this algebra is further extended by the
presence of symmetry, e.g. Lorentz generators, which together
with the position and momentum generators form the so-called extended
spacetime-Poincar\'{e} algebra (also called the extended phase space, see e.g.
\cite{sigma6} and the references therein). Various applications to the
description of Quantum Gravity effects rely on a suitable (quantum)
deformation of both a coordinate algebra as well as a corresponding symmetry (Hopf) algebra
(see e.g. \cite{grb}). In such cases besides the traditional position-momentum noncommutativity
one postulates following \cite{Snyder,DFR95} also the
noncommutativity between position variables and/or, less frequently, between momentum variables.
Such kind of theories can be also considered as noncommutative versions of Quantum Mechanics \cite{nc_qm,Toppan}.
Drinfeld twist techniques turn out to be useful tool in their construction. This point will be a subject of the present note.

\subsection{Quasitriangular Hopf Algebras \protect\cite{Majid90, Majid-book}
and Drinfeld twist techniques \protect\cite{Drinfeld}}

\label{triang}

Let $(\mathcal{H},R)$ be a quasi-triangular bialgebra with the universal
quantum R-matrix $R=R_{1}\otimes R_{2}\in H\otimes H$ satisfying
\begin{equation}  \label{qt}
R\Delta (X)R^{-1}=\Delta ^{op}(X),\quad (\Delta \otimes id){R}={R}_{13}{R}%
_{23},\quad (id\otimes \Delta ){R}={R}_{13}{R}_{12},\quad (\epsilon \otimes
id){R}=(id\otimes \epsilon ){R}=1
\end{equation}
which imply quantum Yang-Baxter equation
\begin{equation}\label{qyb}
  R_{12}R_{13}R_{23} = R_{23}R_{13}R_{12}
\end{equation}
As is  known \cite{Yetter},\cite {Lambe}  any (left) module $A$ over $(\mathcal{H},R)$ becomes
automatically a (left-right) Yetter-Drinfeld module   with the right coaction $\delta _{R}(a)=(R_{2}\rhd a)\otimes R_{1}$
for all $a\in A$.\footnote{%
Throughout the paper we shall be using a shorthand notation of Sweedler
type. The coproduct is denoted as $\Delta (L)=L_{(1)}\otimes L_{(2)}$. For
elements $R\in H\otimes H$ we write $R=R_{1}\otimes R_{2}$,
If $R$ is invertible we write $R^{-1}=\bar{R}_{1}\otimes \bar{R}_{2}$:
$\bar{R}_{1}R_{1^{\prime}}\otimes \bar{R}_{2}R_{2^{\prime }}=\bar{R}_{1^{\prime }}R_{1}\otimes \bar{R%
}_{2^{\prime }}R_{2}=1_{H}\otimes 1_{H}$.}

The category of all (left-right) Yetter-Drinfeld modules $_{\mathcal{H}}%
\mathfrak{YD}^{\mathcal{H}}$ is a prebraided (and braided if $\mathcal{H}$ is a Hopf algebra) monoidal category \cite{Radford}. In particular, a left $H$-
module algebra $\mathcal{A}=(A,\star ,1_{A})$ is an algebra in $_{\mathcal{H}%
}\mathfrak{YD}^{\mathcal{H}}$ if and only if it is a braided commutative,
i.e.
\begin{equation}
a\star b=(R_{2}\rhd b)\star (R_{1}\rhd a)  \label{bcom}
\end{equation}


Let $F\in H\otimes H$ be a normalized cocycle twist in $(\mathcal{H},R)$, i.e an invertible element which satisfies the following conditions
\begin{equation}\label{coc}
F_{12}(\Delta \otimes id)\left( F\right) =F_{23}(id\otimes \Delta )\left(
F\right) \Leftrightarrow F_{1^{\prime }}\left( F_{1}\right) _{\left(
1\right) }\otimes F_{2^{\prime }}\left( F_{1}\right) _{\left( 2\right)
}\otimes F_{2}=F_{1}\otimes F_{1^{\prime }}\left( F_{2}\right) _{\left(
1\right) }\otimes F_{2^{\prime }}\left( F_{2}\right) _{\left( 2\right) }
\end{equation}%
\begin{equation}\label{nor}
\left( \epsilon \otimes id\right)( F)=1_{H}\otimes 1_{H}=(id\otimes \epsilon (F)
\end{equation}%
Its inverse satisfies the similar conditions (according to our  notation
$F=F_{1}\otimes F_{2}\ ,\qquad F^{-1}=\bar{F}_{1}\otimes \bar{F}_{2}$):
\begin{equation}
\left( (\Delta \otimes id)F^{-1}\right) F_{12}^{-1}=\left( (id\otimes \Delta
)F^{-1}\right) F_{23}^{-1}\Leftrightarrow \left( \bar{F}_{1}\right) _{\left(
1\right) }\bar{F}_{1^{\prime }}\otimes \left( \bar{F}_{1}\right) _{\left(
2\right) }\bar{F}_{2^{\prime }}\otimes \bar{F}_{2}=\bar{F}_{1}\otimes \left(
\bar{F}_{2}\right) _{\left( 1\right) }\bar{F}_{1'}\otimes \left( \bar{F}%
_{2}\right) _{\left( 2\right) }\bar{F}_{2'}  \label{coc1}
\end{equation}%
\begin{equation}
\left( \epsilon \otimes id\right) F^{-1}=1_{H}\otimes 1_{H}=(id\otimes
\epsilon )F^{-1}  \label{nor1}
\end{equation}
The twisting element serves the purpose of deformation both the bialgebra structure $\Delta\mapsto\Delta^F=F\Delta F^{-1}$ as well the corresponding module algebra structure $\star\mapsto\star_F=\star\circ(\bar F_1\triangleright\otimes \bar F_2\triangleright )$. We shall denote these new algebras as $\H^F=(H, \Delta^F, \epsilon)$ and $\A_F=(A, \star_F)$.
Moreover, $(\mathcal{H}^{F},R^{F}\equiv F_{21}RF^{-1})$ is quasi-triangular
and the module algebra $(A_{},\star _{F})\in \,_{\mathcal{H}%
^{F}}\mathfrak{YD}^{\mathcal{H}^{F}}$ if and only if $(A,\star )\in \,_{%
\mathcal{H}}\mathfrak{YD}^{\mathcal{H}}$. Drinfeld twisting techniques are very useful in mathematical physics, noncommutative geometry (see e.g \cite{Paolo}) when looking for new quantum spaces and their quantum symmetries (e.g. \cite{sigma10}).

\subsection{Bialgebroids \protect\cite{Lu,Xu,BS}}

The bialgebroid $\mathcal{M}=(M,A,s,t,\Delta,\epsilon)$ consists of a total algebra $M$ and a base
algebra $A$
and the following data:

B1) Two mappings: an algebra homomorphism $s:A\shortrightarrow M$ called a
source map and an algebra anti-homomorphism $t:A\shortrightarrow M$ called a
target map such that: $s(a)t(b)=t(b)s(a)$ is satisfied for all $a,b\in A$.

We consider a left bialgebroid $\mathcal{M}$ as an $A-$bimodule
(with the bimodule structure which prefers the left side) as follows: $%
a.m.b=s\left( a\right) t\left( b\right) m$ for all $a,b\in A$, $m\in M$.

B2) Additionally, it is equipped with coproduct and counit maps. Coproduct and counit make $M$ an $A$-coring \cite{BM} (with axioms like that of a coalgebra such that all mappings are $A$-bimodule homomorphisms and all tensors are over $A$).

More exactly, the bialgebroid coproduct map $\Delta:M\rightarrow M\otimes _A%
M $ is an $A$-bimodule map, where $M\otimes _A M$ is constructed in such a way that $(t\left( a\right) m)\otimes _A n=m_{}\otimes _A(s\left( a\right) n)$; simplifying  the notation one can write $(m.a)\otimes_A\, n= m\otimes_A \,(a.n) $. This is due to the fact that as an Abelian (additive) group  $M\otimes _A M$ is a quotient group of $M\otimes M$ by a subgroup generated by the elements
$\{(t\left( a\right)\otimes 1-1\otimes s\left( a\right))m\otimes n :a\in A, m,n\in M\}$. This subgroup is, in fact, a left ideal in the algebra $M\otimes M$.   However,
$M\otimes _A M$ (unlike $M\otimes M$)is not an algebra in general. To fix this problem one introduces the so-called Takeuchi product $M\times _A M$ \cite{Takeuchi}.\footnote{We follow notational convention introduced in the previous section also for elements of $M\otimes_A M$.}
It is defined as a subgroup of invariant elements  $M\times _A M= \{ m\otimes_A n\in  M\otimes _A M: (m t\left( a\right))\otimes_A n =m\otimes_A n(s\left( a\right)); \forall a\in A \}$
which has natural (component-wise) multiplication ($(m\otimes_A n)(p\otimes_A q)=m p\otimes_A n q$).
Both $M\otimes _A M$ and $M\times _A M$ inherit $A$-bimodule structure determined by the action $m\otimes_A n\mapsto (s(a)m)\otimes_A (t(b)n)$, or $a.(m\otimes_A n).b=(a.m)\otimes_A (n.b)$.
Now we can request additionally that  the image of the coproduct map is in $M\times_A M$,  i.e. that one deals, in fact, with the algebra map: $\Delta(m n)=\Delta(m) \Delta(n)\equiv m_{(1)}n_{(1)}\otimes_A m_{(2)}n_{(2)}$.

The counit map $\epsilon : M\shortrightarrow A$ has to satisfy:
\begin{equation}\label{eps}
  \epsilon(1_{M})=1_{A}, \qquad\quad \epsilon(m n)
=\epsilon(ms(\epsilon (n)))=\epsilon(m t(\epsilon (n))), \quad s(\epsilon(m_{(1)}))m_{(2)}=t(\epsilon(m_{(2)}))m_{(1)}=m
\end{equation}
The axioms \cite{Lu} are similar to those of a bialgebra but are complicated by the
possibility that $A$ is a noncommutative algebra, instead of a commutative ring $K$,
or its images under $s$ and $t$ are not in the center of $M$.

In a case of Hopf algebroids, one additionally assumes that an antipode $\tau: M\shortrightarrow M$ is  to be an algebra
anti-automorphism satisfying conditions of exchanging the source and target
maps and satisfying two axioms similar to the  Hopf algebra antipode axioms. The different versions of introducing the antipode map are possible (see e.g. the second reference in \cite{BS}).

We recall that a morphism between two bialgebroids: $(M,
A,s,t,\Delta ,\epsilon )$ and $(M^{\prime },A,s^{\prime },t^{\prime },\Delta ^{\prime },\epsilon ^{\prime })$ over the
same algebra $A$ consists of algebra map $\phi :M%
\rightarrow M^{\prime }$ such that $\phi \circ s=s^{\prime }$, $%
\phi \circ t=t^{\prime }$, $\epsilon =\epsilon ^{\prime }\circ \phi $ and
the following diagram commutes
\begin{equation}
\begin{array}[t]{ccc}
M & \xrightarrow{\ \ \ \phi\ \ \ } & M^{\prime } \\
&  &  \\
\Delta \downarrow &  & \downarrow \Delta ^{\prime } \\
&  &  \\
M\otimes _{A}M & \xrightarrow{\ \phi
\otimes_A \phi\ } & M^{\prime }\otimes _{A}M
^{\prime } \\
&  &
\end{array}
\label{diag}
\end{equation}%
i.e. $\Delta ^{\prime }\circ \phi =(\phi \otimes _{A}\phi )\circ
\Delta $. For the case of Hopf algebroids with an antipode $\tau$ one should also assume that $\tau'\circ \phi=\phi\circ\tau$.

\subsection{Smash product algebras as bialgebroids \protect\cite{BM}}

In \cite{BM} it was shown how a Hopf algebroid structure can be associated to
a smash product of a Hopf algebra with a braided commutative algebra in
the Yetter-Drinfeld $\mathfrak{YD}$ category.
The theorem [Theorem 4.1 in \cite{BM}] adapted to our needs reads as follows:

\begin{theorem}
\label{thBM}
Let $\H=(H, \Delta, \epsilon)$ be a bialgebra, $\A=\left( A,\star \right) $ is a left $\H$-module
algebra and $\left( \A,\rho\right) $ a right $\H$-comodule~.

If  $\left( A,\star ,\rho\right) $\footnote{$%
\rho$ is an algebra map called coaction: $\rho : A\rightarrow
A\otimes H^{op},\ \rho\left( a\right) =a_{< 0> }\otimes
a_{< 1>}$~in Sweedler notation.} is a
braided commutative algebra in $_{\H}\mathfrak{YD}^{\H}$   then $%
\left( \A\rtimes \H,s,t,\tilde\Delta,\tilde\epsilon \right) $ is an $\A$-bialgebroid
with the source, target, coproduct and the counit given by the following maps:
\begin{eqnarray}
s\left( a\right)  &=&a\rtimes 1_{H},\quad t\left( a\right)\equiv\rho(a) =a_{<
0> }\rtimes a_{< 1> }  \label{st} \\
\tilde\Delta ^{}\left( a\rtimes L\right)  &=&(a\rtimes L_{\left( 1\right) })\otimes
_{\A}(1_{A}\rtimes L_{\left( 2\right) })  \label{cop} \\
\tilde\epsilon \left( a\rtimes L\right)  &=&\epsilon _{}(L)a  \label{coun}
\end{eqnarray}
for all $a\in A$ and $L\in H$.
\end{theorem}
Thus $(a\rtimes L)\blacktriangleright b=a\star (L\triangleright b)$. In particular, $(a\rtimes 1_H)\blacktriangleright b=a\star b$ acts by multiplication from the left, while $(1_A\rtimes L)\blacktriangleright b= L\triangleright b$ preserves the initial action.

\subsection{Twisted bialgebroids \cite{Xu}}

The category of bialgebroids (Hopf algebroids) was introduced by P. Xu in \cite{Xu} \footnote{It should be noted that Xu's paper concerns, in fact, bialgebroids (his definition does not include the antipode map).}.
Let us recall the relevant results from \cite{Xu} on twist deformation of bialgebroids. Before proceeding further one should remember that bialgebroid definition provides a canonical action $\blacktriangleright: M\otimes A\rightarrow A$:
\footnote{This action is sometimes  referred as an anchor  $M\ni m\rightarrow m\blacktriangleright\in End A$, see \cite{BM}.}
 \begin{equation}\label{eps1}
m\blacktriangleright a= \epsilon(ms(a))=\epsilon(m t(a)),
\end{equation}
induced by the counit $\epsilon$ (cf. (\ref{eps})).
We should point out that the multiplication in $M$ does not change like in the case of Drinfeld theory, where twist deformation modifies
coalgebraic sector only. Nevertheless  the one in $A$ changes ($A\mapsto A_F$):
$$\cdotp\, \mapsto\, \cdotp_F=\cdotp\circ(\bar F_1\blacktriangleright\otimes \bar F_2\blacktriangleright)\,.$$
We are now in position to present the simplified version of [Theorem. 4.14 in \cite{Xu}], skipping some details which are not relevant for our considerations (following Drinfeld convention our twist is inverse with respect to the one considered by Xu in \cite{Xu}.):
\begin{theorem}\label{thXu}
Assume that $\left( M,A,s,t,\Delta,\epsilon \right) $ is
bialgebroid over the algebra $A$ and $F=F_1\otimes_A F_2\in M\otimes _{A}M$ is a "twistor" (Hopf algebroid twist \footnote{It satisfies the same Drinfeld conditions (\ref{coc})-(\ref{nor1}) with $\otimes$ replaced by $\otimes_A$.}).
Then $\left( M, A_{F}, s_{F}, t_{F}, \Delta_{F}, \epsilon\right) $ is a bialgebroid over the algebra $A_F$, where
\begin{equation}
s_{F}(a)=s\left( \bar{F}_{1}\blacktriangleright a\right) \bar{F}_{2}\quad
;\quad \ t_{F}(a)=t\left( \bar{F}_{2}\blacktriangleright a\right) \bar{F}%
_{1}\quad \ \forall a\in A.  \label{stXu}
\end{equation}%
and new twisted coproduct $\Delta_{F}:M\shortrightarrow M\otimes
_{A_{F}}M$ :
\begin{equation}
\Delta_{F}\left( m\right) =F^\#\left(\Delta\left(
m\right) F^{-1}\right),\quad \forall \ \ m\in M  \label{copXu}
\end{equation}%
\end{theorem}

The map $F^\#: M\otimes_A M\rightarrow M\otimes_{A_F} M$ is defined by (cf. Corollary 4.4  in \cite{Xu}):
\begin{equation}\label{hash}
F^\#(m\otimes_A n)= (F_1\, m)\otimes_{A_F} (F_2\, n).
\end{equation}

\section{Main result}

In this section we are going to revisit
bialgebroids in a context of twist deformation of smash product algebras
. It appears according to construction of
Brzezinski-Militaru from \cite{BM} that the smash product algebras, under suitable
assumptions, can be equipped with the bialgebroid structures. Moreover, we are
going to show that the  bialgebroid obtained by bialgebroid twisting \cite{Xu} of
the smash product algebra and bialgebroid obtained from the smash
product algebra of twisted bialgebra with its twisted module algebra are equivalent
(isomorphic).

Let $\mathcal{H}\equiv \left( H,\Delta ,\epsilon \right) $ be a bialgebra
and $\mathcal{A}\equiv (A,\star )$ be a (left) module algebra over $\mathcal{%
H}$. We denote the corresponding (left) action as $\triangleright :H\otimes
A\rightarrow A$: $L\triangleright (a\star b)=(L_{(1)}\triangleright a)\star
(L_{(2)}\triangleright b)$. Assume that $F=F_{1}\otimes F_{2}\in H\otimes H$
is a normalized cocycle twist for $\mathcal{H}$. It allows us to construct
new bialgebra $\mathcal{H}^{F}\equiv \left( H,\Delta ^{F},\epsilon \right) $
and new module algebra $\mathcal{A}_{F}\equiv (A,\star _{F})$ with the same
action, where $\Delta ^{F}=F\Delta F^{-1}$ and $\star _{F}=\star
(F^{-1}\circ (\triangleright \otimes \triangleright ))$ with $F^{-1}=\bar{F}%
_{1}\otimes \bar{F}_{2}$. Actually, as far as smash product is concerned,
it turns out that:
\begin{proposition}
For any Drinfel'd twist $F$ two smash product algebras $%
\mathcal{A}\rtimes \mathcal{H}$ and $\mathcal{A}_{F}\rtimes \mathcal{H}^{F}$
are isomorphic, even though the algebras $\mathcal{A}$ and $\mathcal{A}^{F}$
are not isomorphic and $\mathcal{H}$ and $\mathcal{H}^{F}$ are not
isomorphic as bialgebras (see e.g. \cite{sigma6}).
\end{proposition}
Although this fact seems to be known, e.g. \cite{Bulacu}, we provide the proof for completeness.\\
Proof:\\
Firstly, we recall that
both algebras are determined on the same $K$-module $A\otimes H$ but differ
by the multiplications (cf. (\ref{cr})):
\begin{equation}
(a\rtimes L)\star (b\rtimes J)=a\star (L_{(1)}\triangleright b)\rtimes
L_{(2)}J\ ,\quad (a\rtimes L)\star _{F}(b\rtimes J)=a\star
_{F}(L_{(1^{F})}\triangleright b)\rtimes L_{(2^{F})}J  \label{m1}
\end{equation}%
where $\Delta ^{F}(L)=F\Delta (L)F^{-1}=L_{(1^{F})}\otimes
L_{(2^{F})}=F_{1}L_{(1)}\bar{F}_{1^{\prime }}\otimes F_{2}L_{(2)}\bar{F}%
_{2^{\prime }}$ is the twisted coproduct of the bialgebra $\mathcal{H}^{F}$.
Both algebras are generated by simpler elements: $a\rtimes 1_{H},a\in A$ and
$1_{A}\rtimes L,L\in H$, i.e. $a\rtimes L=(a\rtimes 1_{H})\star
(1_{A}\rtimes L)=(a\rtimes 1_{H})\star _{F}(1_{A}\rtimes L)$. Of course, the unit $1_A\rtimes 1_H$ is the same
for both multiplications.

The isomorphism $\varphi :\mathcal{A}_{F}\rtimes \mathcal{H}^{F}\rightarrow
\mathcal{A}\rtimes \mathcal{H}$ can be defined by the formula
\begin{equation}
\varphi \left( a\rtimes L\right) =\left( \bar{F}_{1}\triangleright a\right)
\rtimes \bar{F}_{2}L  \label{def_phi}
\end{equation}%
such that:
\begin{equation}
\varphi \left( (a\rtimes L)\star _{F}(b\rtimes J)\right) =\varphi \left(
a\rtimes L\right) \star \varphi \left( b\rtimes J\right)   \label{varp}
\end{equation}%
for all $a,b\in A$ and $L,J\in H$. One notices that due to the normalization
condition $\varphi (1_{A}\rtimes L)=1_{A}\rtimes L$. The inverse map $%
\varphi ^{-1}:\mathcal{A}\rtimes \mathcal{H}\rightarrow \mathcal{A}%
_{F}\rtimes \mathcal{H}^{F}$ is, of course, given by $\varphi ^{-1}\left(
a\rtimes L\right) =\left( F_{1}\triangleright a\right) \rtimes F_{2}L$.

We begin by checking  the equality (\ref{varp}) for some special cases. Firstly we take\medskip\\

i) $\varphi \left( (a\rtimes 1_H)\star _{F}(b\rtimes J)\right) =\varphi \left(
a\rtimes 1\right) \star \varphi \left( b\rtimes J\right) $.

In $\mathcal{A}_{F}\rtimes \mathcal{H}^{F}$:\quad $(a\rtimes
1)\star _{F}(b\rtimes J)=\left( a\star _{F}b\right) \rtimes J$ and in $%
\mathcal{A}\rtimes \mathcal{H}$:\quad $(a\rtimes 1_H)\star (b\rtimes J)=\left(
a\star b\right) \rtimes J$.

On one hand, from the above and from (\ref{varp}) we have the following series of equalities:
\begin{eqnarray*}
\varphi \left( (a\rtimes 1_H)\star _{F}(b\rtimes J)\right)  &=&\varphi \left(
\left( a\star _{F}b\right) \rtimes J\right) = \\
\left( \bar{F}_{1}\triangleright (a\star _{F}b)\right) \rtimes \bar{F}_{2}J
&=&\bar{F}_{1}\triangleright \left[ \left( \bar{F}_{1^{\prime
}}\triangleright a\right) \star \left( \bar{F}_{2^{\prime }}\triangleright
b\right) \right] \otimes \bar{F}_{2}J= \\
&=&\left[ (\bar{F}_{1})_{(1)}\bar{F}_{1^{\prime }}\triangleright a\right]
\star \left[ (\bar{F}_{1})_{(2)}\bar{F}_{2^{\prime }}\triangleright b\right]
\rtimes \bar{F}_{2}J= \\
&=&\left( \bar{F}_{1}\triangleright a\right) \star \left( \left( \bar{F}%
_{2}\right) _{\left( 1\right) }\bar{F}_{1'}\triangleright b\right) \rtimes
\left( \bar{F}_{2}\right) _{\left( 2\right) }\bar{F}_{2'}J
\end{eqnarray*}
where we used the cocycle identity for the inverse twist (\ref{coc1}).

On the other hand we have from (\ref{varp}) and the above:
\begin{eqnarray*}
\varphi \left( a\rtimes 1_H\right) \star \varphi \left( b\rtimes J\right)
&=&[\left( \bar{F}_{1}\triangleright a\right) \rtimes \bar{F}_{2}]\star
\lbrack \left( \bar{F}_{1^{\prime }}\triangleright b\right) \rtimes \bar{F}%
_{2^{\prime }}J]= \\
&=&\left( \bar{F}_{1}\triangleright a\right) \star \left( \left( \bar{F}%
_{2}\right) _{\left( 1\right) } \bar{F}_{1^{\prime }}\triangleright b \right)
\rtimes \left( \bar{F}_{2}\right) _{\left( 2\right) }\bar{F}_{2^{\prime }}J=
\\
&=&\varphi \left( (a\rtimes 1_H)\star _{F}(b\rtimes J)\right)
\end{eqnarray*}

As a next step we consider\\
ii) $\varphi \left( (1_A\rtimes L)\star _{F}(b\rtimes J)\right) =\varphi
\left( 1_A\rtimes L\right) \star \varphi \left( b\rtimes J\right)$

In $\mathcal{A}_{F}\rtimes \mathcal{H}^{F}$ one calculates: $$(1_A\rtimes L)\star_{F}(b\rtimes J)=
\left( L_{(1^{F})}\triangleright b\right) \rtimes L_{(2^{F})}J=
\left( F_1 L_{(1^{})}\bar{F}_{1'}\triangleright b\right) \rtimes F_2 L_{(2^{})}\bar F_{2'}J=
\varphi^{-1}(\left(L_{(1^{})}\bar{F}_{1'}\triangleright b\right) \rtimes L_{(2^{})}\bar F_{2'}J)$$
where the first equality is due to the normalization condition $\epsilon \left( \bar{F}_{1}\right)\bar{F}_{2}=1_H=%
\bar{F}_{1}\epsilon \left( \bar{F}_{2}\right)$.
Therefore one has the following equalities:
\begin{eqnarray*}
\varphi \left( (1\rtimes L)\star _{F}(b\rtimes J)\right)  &=&
\left(L_{(1^{})}\bar{F}_{1'}\triangleright b\right) \rtimes L_{(2^{})}\bar F_{2'}J=
(1_A\rtimes L)\star(\left(\bar{F}_{1'}\triangleright b\right) \rtimes \bar F_{2'}J)\\
&=&\varphi(1_A\rtimes L)\star\varphi( b \rtimes J)
\end{eqnarray*}

Verification on the remaining pair of generators: $(a\rtimes 1_H)\star_F (1_A\rtimes J)=a\rtimes J$ and
$(1_A\rtimes L)\star_F (1_A\rtimes J)=1_A\rtimes LJ$ is rather straightforward. \smallskip\\

To finish the proof one checks (\ref{varp}) using above partial results:
\begin{eqnarray*}
\varphi \left( (a\rtimes L)\star _{F}(b\rtimes J)\right)  &=&\varphi \left( (a\rtimes 1_H)\star _{F}((\tilde L_{(1^{})}\triangleright b)\rtimes \tilde L_{(2)}J)\right)
 =\varphi \left( a\rtimes 1_H\right)\star _{}\varphi\left((\tilde L_{(1^{})}\triangleright b)\rtimes \tilde L_{(2)}J\right)
 \\
&=&\varphi \left( a\rtimes 1_H\right)\star _{}\varphi\left((1_A\rtimes L)\star_F(b\rtimes J)\right)
=\varphi(a\rtimes 1_H)\star\varphi(1_A\rtimes L)\star\varphi( b \rtimes J)\\
&=&\varphi(a\rtimes L)\star\varphi( b \rtimes J)
\end{eqnarray*}
The proof is done.\medskip\\

From now on we assume that $(\mathcal{H},R)$ is quasi-triangular bialgebra
and the algebra $\mathcal{A}$ is braided commutative in the category $_{%
\mathcal{H}}\mathfrak{YD}^{\mathcal{H}}$, i.e. $a\star
b=(R_{2}\triangleright b)\star (R_{1}\triangleright a)$. Then according to Brzezinski-Militaru
construction (Theorem \ref{thBM}, \cite{BM}) $\mathcal{A}%
\rtimes \mathcal{H}$ is a bialgebroid over the algebra $\mathcal{A}$ if we
define (shifting $\Delta :H\rightarrow H\otimes H$ to $\tilde{\Delta}:%
\mathcal{A}\rtimes \mathcal{H}\rightarrow (\mathcal{A}\rtimes \mathcal{H}%
)\otimes _{\mathcal{A}}(\mathcal{A}\rtimes \mathcal{H})$)
\begin{equation}
\tilde{\Delta}(a\rtimes L)=(a\rtimes L_{(1)})\otimes _{\mathcal{A}%
}(1_{A}\rtimes L_{(2)}),\quad s\left( a\right) =a\rtimes 1_{H},\quad t\left(
a\right) =\left( R_{2}\triangleright a\right) \rtimes R_{1},\quad \tilde{%
\epsilon}(a\rtimes L)=\epsilon (L)a  \label{bm1}
\end{equation}%
The bialgebroid counit map $\tilde{\epsilon}$ will not change through the
rest of this note. Following the same idea we can shift the quantum R-matrix
from the bialgebra $\H$ to the bialgebroid $\A\rtimes\H$
\begin{equation}  \label{bm1a}
R\rightarrow\tilde R=(1_A\rtimes R_{1})\otimes _{\mathcal{A}}(1_A\rtimes
R_{2})\in (\mathcal{A}\rtimes \mathcal{H})\otimes_\mathcal{A} (\mathcal{A}%
\rtimes \mathcal{H}).
\end{equation}
One can easily check by direct calculations that properties of the cocycle
type will be preserved (cf. (\ref{qt})):
\begin{equation}  \label{qt1}
(\tilde\Delta \otimes_\A id){\tilde R}={\tilde R}_{13}{\tilde R}_{23},\quad
(id\otimes_\A \tilde\Delta ){\tilde R}={\tilde R}_{13}{\tilde R}_{12},\quad
(\tilde\epsilon \otimes_\A id){\tilde R}=(id\otimes_\A \tilde\epsilon ){\tilde R}%
=(1_A\rtimes 1_H)\otimes_\mathcal{A} (1_A\rtimes 1_H)
\end{equation}
while the remaining is lost
\begin{equation}  \label{qt2}
\tilde R\tilde\Delta (a\rtimes L)\tilde R^{-1}=((R_1\triangleright a)\rtimes
L_{(2)})\otimes_\mathcal{A} (1_A\rtimes R_2L_{(1)})\neq \tilde\Delta
^{op}(a\rtimes L) = ((R_2\triangleright a)\rtimes R_1L_{(2)})\otimes_%
\mathcal{A} (1_A\rtimes L_{(1)}),
\end{equation}
where in the last equation we have used the property $X\otimes_\mathcal{A}
s(a)Y=t(a)X\otimes_\mathcal{A} Y$. Therefore, any Drinfeld twist $%
F=F_{1}\otimes F_{2}\in H\otimes H$ in the bialgebra $\mathcal{H}$ can be
also shifted to the bialgebroid twist $\tilde{F}\in (\mathcal{A}\rtimes
\mathcal{H})\otimes _{\mathcal{A}}(\mathcal{A}\rtimes \mathcal{H})$ by
\footnote{%
Our twist is inverse with respect to the one considered by Xu in \cite{Xu}.}
\begin{equation}\label{qt3}
F\rightarrow\tilde{F}=(1_A\rtimes F_{1})\otimes _{\mathcal{A}}(1_A\rtimes
F_{2})
\end{equation}%
which automatically satisfies bialgebroid cocycle and normalization
conditions.

Similarly, the construction of Brzezinski-Militaru (Theorem \ref{thBM}, \cite{BM}%
) makes $\mathcal{A}_{F}\rtimes \mathcal{H}^{F}$ a bialgebroid over the
algebra $\mathcal{A}_{F}$ if we set
\begin{equation}
\widetilde{\Delta^{F}}(a\rtimes L)=(a\rtimes L_{(1^{F})})\otimes _{\mathcal{A%
}_{F}}(1_{A}\rtimes L_{(2^{F})}),\quad s^{F}\left( a\right) =a\rtimes
1_{H},\quad t^{F}\left( a\right) =\left( R_{2}^{F}\triangleright a\right)
\rtimes R_{1}^{F}  \label{bm3}
\end{equation}%
where $R^{F}=F_{21}RF^{-1}=F_{2^{\prime }}R_{1}\bar{F}_{1^{\prime \prime
}}\otimes F_{1^{\prime }}R_{2}\bar{F}_{2^{\prime \prime }}$ and the algebra $%
\mathcal{A}_{F}$ is braided commutative as well: $a\star
_{F}b=(R_{2}^{F}\triangleright b)\star _{F}(R_{1}^{F}\triangleright a)$.
More explicitly
\begin{equation}  \label{bm3a}
\widetilde{\Delta^{F}}(a\rtimes L)=(a\rtimes F_1 L_{(1)}\bar F_{1^{\prime
}})\otimes _{\mathcal{A}_{F}}(1_{A}\rtimes F_2 L_{(2)}\bar F_{2^{\prime }})
\end{equation}

As a next task, according to Xu (Theorem \ref{thXu}), one applies bialgebroid
twisting in order to obtain new twisted bialgebroid $(\mathcal{A}\rtimes
\mathcal{H})^{\tilde{F}}$ by making use of the twist (\ref{qt3}):
\begin{equation}
\tilde{\Delta}_{\tilde{F}}(a\rtimes J)=\tilde{F}^{\#}(\tilde{\Delta}%
(a\rtimes J)\tilde{F}^{-1}),\quad s_{\tilde{F}}\left( a\right) =(\bar{F}%
_{1}\triangleright a)\rtimes \bar{F}_{2},\quad t_{\tilde{F}}\left( a\right)
=\left( R_{2}\bar{F}_{2^{\prime }}\triangleright a\right) \rtimes R_{1}\bar{F%
}_{1^{\prime }}  \label{bm2}
\end{equation}%
where, in our case, $\tilde{F}^{\#}:(\mathcal{A}\rtimes \mathcal{H})\otimes _{\mathcal{A}}(%
\mathcal{A}\rtimes \mathcal{H})\rightarrow (\mathcal{A}\rtimes \mathcal{H}%
)\otimes _{\mathcal{A}_{F}}(\mathcal{A}\rtimes \mathcal{H})$ is determined by the formula (cf.(\ref{hash}))
\begin{equation}  \label{bm2a}
\tilde{F}^{\#}((a\rtimes L)\otimes _{\mathcal{A}}(b\rtimes
J))=(((F_{1})_{(1)}\triangleright a)\rtimes (F_{1})_{(2)}L)\otimes _{%
\mathcal{A}_{F}}(((F_{2})_{(1)}\triangleright b)\rtimes (F_{2})_{(2)}J)
\end{equation}
Since $(1_A\rtimes M)\blacktriangleright a=M\triangleright a$ for any $M\in H$ the base algebra is just $\A_F$.

Our goal is to compare bialgebroids $\mathcal{A}_{F}\rtimes \mathcal{H}^{F}$
and $(\mathcal{A}\rtimes \mathcal{H})^{\tilde{F}}$. In fact we are going to
prove the following

\begin{theorem}
Let $(\mathcal{H}, R)$ be a quasi-triangular bialgebra and $\mathcal{A}$
stands for braided commutative module algebra w.r.t. $(\mathcal{H}, R)$.
Assume that $F=F_1\otimes F_2\in H\otimes H$ is a normalized cocycle twist in $\mathcal{H}$. Then
\begin{equation}  \label{isom}
\mathcal{A}_{F}\rtimes \mathcal{H}^{F}\cong \left( \mathcal{A}\rtimes
\mathcal{H}\right) ^{\tilde{F}}
\end{equation}
are isomorphic bialgebroids, where $\tilde F$ denotes bialgebroid cocycle twist (\ref{qt3}) obtained from $F$.
\end{theorem}

Proof:

All properties of bialgebroid are fulfilled according to Theorems \ref{thBM} and \ref{thXu}. The base algebra $\mathcal{A}_{F}$ is the same on both sides. Therefore we can use the isomorphism (\ref{varp}) $\varphi :%
\mathcal{A}_{F}\rtimes \mathcal{H}^{F}\rightarrow \mathcal{A}_{{}}\rtimes
\mathcal{H}^{{}}$ of total algebras and at the same time, demonstrate that the
following diagram commutes (cf. (\ref{diag}))
\begin{equation}
\begin{array}[t]{ccc}
\mathcal{A}_{F}\rtimes \mathcal{H}^{F} & \xrightarrow{\ \ \ \ \varphi\ \ \ \
\ } & \mathcal{A}_{{}}\rtimes \mathcal{H}^{{}} \\
&  &  \\
\widetilde{\Delta^{F}} \downarrow &  & \downarrow \tilde{\Delta}_{\tilde{F}%
}^{{}} \\
&  &  \\
(\mathcal{A}_{F}\rtimes \mathcal{H}^{F})\otimes _{\mathcal{A}_{F}}(\mathcal{A%
}_{F}\rtimes \mathcal{H}^{F}) & \xrightarrow{\ \ \varphi \otimes_{\A^F}
\varphi\ \ } & (\mathcal{A}_{{}}\rtimes \mathcal{H})\otimes _{\mathcal{A}%
_{F}}(\mathcal{A}_{{}}\rtimes \mathcal{H}) \\
&  &
\end{array}
\label{diag2}
\end{equation}%
i.e. $\tilde{\Delta}_{\tilde F}\circ \varphi =(\varphi \otimes _{\mathcal{A}%
_{F}}\varphi )\circ \widetilde{\Delta^{F} }$.\newline

The coproduct $\tilde{\Delta}_{\tilde{F}}$ can be found in more explicit
form as
\begin{equation}  \label{bm2b}
\tilde{\Delta}_{\tilde{F}}(a\rtimes J)=(((F_1)_{(1)}\triangleright a)\rtimes
(F_1)_{(2)}J_{(1)}\bar F_{1^{\prime }}) \otimes_{\mathcal{A}_F} ( 1_A\rtimes
F_2J_{(2)}\bar F_{2^{\prime }})
\end{equation}
since $\epsilon((F_2)_{(1)})(F_2)_{(2)}=F_2$.

In order to simplify the proof we check the diagram (\ref{diag2}) on
generators. We begin from
\begin{equation*}
\tilde{\Delta}_{\tilde F}(\varphi(1_A\rtimes J))=\tilde{\Delta}_{\tilde
F}(1_A\rtimes J)=  (((F_1)_{(1)}\triangleright 1_A)\rtimes
(F_1)_{(2)}J_{(1)}\bar F_{1^{\prime }}) \otimes_{\mathcal{A}_F} ( 1_A\rtimes
F_2J_{(2)}\bar F_{2^{\prime }})
\end{equation*}
\begin{equation*}
=(1_A\rtimes F_1J_{(1)}\bar F_{1^{\prime }})\otimes_{\mathcal{A}_F} (
1_A\rtimes F_2J_{(2)}\bar F_{2^{\prime }})=\widetilde{\Delta^{F}}(1_A\rtimes
J)
\end{equation*}

Next we check
\begin{equation*}
\tilde{\Delta}_{\tilde F}(\varphi(a\rtimes 1_H))=\tilde{\Delta}_{\tilde
F}((\bar F_1\triangleright a)\rtimes \bar F_2)= \tilde F^\#\left( ((\bar
F_1\triangleright a)\rtimes (\bar F_2)_{(1)}\bar F_{1^{\prime }}) \otimes_{%
\mathcal{A}} ( 1_A\rtimes (\bar F_2)_{(2)}\bar F_{2^{\prime }})\right)=
\end{equation*}
\begin{equation*}
\tilde F^\#\left( (((\bar F_1)_{(1)}\bar F_{1^{\prime }}\triangleright
a)\rtimes (\bar F_1)_{(2)}\bar F_{2^{\prime }}) \otimes_{\mathcal{A}} (
1_A\rtimes \bar F_2)\right)=((((F_{1^{\prime \prime }})_{(1)}\bar
F_1)_{(1)}\bar F_{1^{\prime }}\triangleright a)\rtimes (F_{1^{\prime \prime
}})_{(2)}(\bar F_1)_{(2)}\bar F_{2^{\prime }})\otimes_{\mathcal{A}_F} (
1_A\rtimes \bar F_{2^{\prime \prime }}F_2)
\end{equation*}
Since $(\Delta\otimes id)(F)(\Delta\otimes id)(F^{-1})=1_H\otimes 1_H$ the
last expression simplifies to
\begin{equation*}
((\bar F_1\triangleright a)\rtimes \bar F_2)\otimes_{\mathcal{A}_F}(1_A\rtimes
1_H)=\varphi(a\rtimes 1_H)\otimes_{\mathcal{A}_F}\varphi(1_A\rtimes 1_H)=
(\varphi \otimes _{\mathcal{A}_{F}}\varphi )\left( \widetilde{\Delta^{F} }%
(a\rtimes 1_H) \right)
\end{equation*}

In order to complete the proof one has to check that $\varphi (s^{F}(a))=s_{%
\tilde{F}}(a)$, $\varphi (t^{F}(a))=t_{\tilde{F}}(a)$ and $\tilde{\epsilon}%
\circ \varphi =\tilde{\epsilon}$. It is not difficult to get these equalities.

For example, for the target maps we have
\begin{equation*}
\varphi (t^{F}(a))=\left( \bar{F}_{1}R_{2}^{F}\triangleright a\right)
\rtimes \bar{F}_{2}R_{1}^{F}=\left( \bar{F}_{1}\left( F_{1^{\prime }}R_{2}%
\bar{F}_{2^{\prime \prime }}\right) \triangleright a\right) \rtimes \bar{F}%
_{2}\left( F_{2^{\prime }}R_{1}\bar{F}_{1^{\prime \prime }}\right) =\left(
R_{2}\bar{F}_{2^{\prime \prime }}\triangleright a\right) \rtimes R_{1}\bar{F}%
_{1^{\prime \prime }}=t_{\tilde{F}}(a)
\end{equation*}
using $R^{F}=R_{1}^{F}\otimes R_{2}^{F}=F_{2^{\prime }}R_{1}\bar{F}%
_{1^{\prime \prime }}\otimes F_{1^{\prime }}R_{2}\bar{F}_{2^{\prime \prime
}} $ and $\bar{F}_{1}F_{1^{\prime }}\otimes \bar{F}_{2}F_{2^{\prime
}}=F^{-1}F=1_H\otimes 1_H$.\\

Therefore the proof is completed.

\subsection{Comments  on crossed product and Hopf-Galois extension}

A smash product $\A\rtimes \H$ is a particular kind of a crossed product algebra $\A\rtimes_\sigma \H$, where a convolution invertible map $\sigma: \H\otimes\H\rightarrow \A$ has to satisfy  (in $\A$) the so-called 2-cocycle
$$[L_{(1)}\rhd   \sigma(J_{(1)}, K_{(1)})]\sigma(L_{(2)}, J_{(2)}K_{(2)})\,=\, \sigma(L_{(1)},J_{(1)})\sigma(L_{(2)}J_{(2)},K), \quad \sigma(J, 1_H)=\sigma(1_H, J)=\epsilon(J) 1_A$$
as well as twisted module
$$[L_{(1)}\rhd (J_{(1)}\rhd a)]\sigma(L_{(2)}, J_{(2)})\,=\, \sigma(L_{(1)},J_{(1)})[(L_{(2)}J_{(2)})\rhd a]$$
conditions for any $a\in\A$ and $L,J,K\in \H$.
These properties allow to establish  on the vector space $\A\otimes \H$ the structure of unital, associative algebra with the multiplication
$$(a\otimes L)(b\otimes J)=a(L_{(1)}\rhd b)\sigma(L_{(2)}, J_{(1)})\otimes L_{(3)}\,J_{(2)}$$
This algebra is denoted as $\A\rtimes_\sigma \H$ \cite{Sus1}.  It has  a natural  left  $\A$ module  and right $\H$ comodule structures (the so-called normal basis property),
which makes it a $H$-comodule algebra (a coring) with the subalgebra $\A\otimes 1_H= (\A\rtimes_\sigma \H)^{coH}$ composed of coinvariants of the coaction.
\footnote{ For a coring $M$ over the Hopf algebra $H$ with the right coaction $\rho: M\rightarrow M\otimes H$ one defines a
subalgebra of coinvariant elements, $M^{coH}=\{h\in M: \rho(m)=m\otimes 1_H\}$. We say that the extension $M^{coH}\subset M$ is $H$-Hopf-Galois if the map
$M\otimes_{M^{coH}}M\rightarrow M\otimes H$, given by $m\otimes n\mapsto (m\otimes 1_H)\rho(n)$, is bijective \cite{Sus1}.
}

Due to this fact  it provides a canonical example of Hopf-Galois extension \cite{Sus1} which, in turn, is an algebraic counterpart of a quantum principal bundle
\cite{Sus1,B-M,ABPS}. Various twist deformations of such principal bundles have been proposed recently in \cite{ABPS}.
Taking the trivial cocycle $\sigma_0(L, J)=\epsilon(L)\epsilon(J)\,1_A$ one reconstructs the smash product. A natural question which appears now is whether the result of the present  section can be extended to the case of nontrivial cocycle $\sigma: \H\otimes\H\rightarrow \A$~?

\section{Lie algebra case}

Nice and simple illustration of the framework presented in this paper is
provided by the Lie algebra $\mathfrak{g}$ itself. It is also important from
physical point of view as the symmetries in physics are described by Lie
algebras. The Lie algebra $\mathfrak{g}$ can be generalized as $U_{\mathfrak{%
g}}$ - universal enveloping algebra to a Hopf algebra (with primitive Hopf
algebra maps) $\mathcal{U}_{\mathfrak{g}}=\left( U_{\mathfrak{g}},\Delta
_{0},\epsilon ,S_{0}\right) $ over the field $\mathbb{K}=\mathbb{C}\,\mbox{or}\, \mathbb{R}$ of complex or real numbers.
Through the deformation procedure (see Sec. \ref{triang}), which requires extension to $%
\mathbb{K}[[h]]$, it becomes $\mathcal{U}_{\mathfrak{g},h}=\left( U_{%
\mathfrak{g}}^{}[[h]],\Delta ^{},\epsilon ,S^{}\right) $ deformed quantum
symmetry algebra (quantum group) of the corresponding noncommutative
quantum spacetime (=Hopf module algebra). In the case of Lie algebras there is well known correspondence between classical and quantum $r$-matrices
\begin{equation}\label{r-m}
  R= 1+ h\, r +O(h^2)
\end{equation}
where $r\in \mathfrak{g}\otimes \mathfrak{g}$ denotes the classical $r$-matrix satisfying, due to (\ref{qyb}), classical Yang-Baxter equation (CYBE). Its skew symmetric part $r-r_{21}$ describes  Poisson-Lie structure on the corresponding Lie group. In fact, there are two types of quantum deformations of Lie algebras: triangular (nonstandard) and quasi-triangular (standard). The former corresponds to the situation when  $r$
is skew-symmetric, i.e. $r\in \mathfrak{g}\wedge \mathfrak{g}$. In this case existing of cocycle twist is ensured by Drinfeld theorem, even if its explicit form not always is known. Knowing the twist $F\in U_\mathfrak{g}^{}[[h]]\otimes U_\mathfrak{g}^{}[[h]]$ one can proceed with the deformation procedure and construct quantum r-matrix $R=F_{21}F^{-1}$ which is triangular. Twisting techniques are well developed and very useful in mathematical physics (see e.g. \cite{Paolo}). In contrast, quasi-triangular deformations, which apply to semi-simple Lie algebras, are related to the classical $r$-matrices with the skew-symmetric part satisfying modified classical Yang-Baxter equation (MCYBE).

Representations of Lie algebra $\mathfrak{g}$  provide examples of module algebras via deformation of  (commutative) algebra of smooth
functions $\mathcal{A}=C^{\infty }(V)$(=algebra of spacetime coordinates) on the corresponding vector space $V$ 
in the following way.
Given  representation $\rho $ induces the action on the vector space $V$:
\begin{equation}
\rho :\mathfrak{g}\rightarrow End_{\mathbb{K}}V\quad \Leftrightarrow \quad
\triangleright :\mathfrak{g}\otimes V\rightarrow V
\end{equation}%
$L\triangleright v\equiv \rho (L)(v)$. This action can be uniquely
extended to the action of the entire universal enveloping algebra $\triangleright
:U_{\mathfrak{g}}\otimes V\rightarrow V$ (and eventually to its topological
extension $U_{\mathfrak{g}}[[h]]$).

Further extension relies on the possibility of replacing $V$ by the
commutative algebra of smooth functions on $V$ in the case of finite dimesional representation. 
Assume $\{e_{i}\}_{1}^{m}$ ($m=dim_{\mathbb{K}}V$) is some basis providing
coordinates for the vectors: $v=x^{i}e_{i}$. Let us denote $\rho
(L)=[L_{i}^{j}]$ as the corresponding matrix. Then we are in position  to
construct the first order differential operators acting on the manifold $V$
or more exactly on the algebra of its smooth functions $C^{\infty }(V)$:
\begin{equation}\label{x1}
\hat{\rho}(L)=-L_{\alpha}^{\beta}x^{\alpha}\partial _{\beta}
\end{equation}%
which is in fact coordinate independent object. Therefore, it defines an extended
action $\triangleright :U_{\mathfrak{g}}\otimes C^{\infty }(V)\rightarrow
C^{\infty }(V)$.\footnote{More generally we can assume that $X$ is a (smooth) $G$-manifold, where $\mathfrak{g}$ denotes Lie algebra of $G$. Then we have the action $U_{\mathfrak{g}}\otimes C^{\infty }(X)\rightarrow C^{\infty }(X)$ provided by so-called Killing vector fields.}

Note that the Leibniz rule
\begin{equation}  \label{leibniz}
\hat{\rho}\left( L\right) \left( a\cdot b\right) =\hat{\rho}\left( L\right)
a\cdot b+a\cdot \hat{\rho}\left( L\right) b
\end{equation}%
for all $a,b\in C^{\infty }(V)$ is automatically satisfied. And since $\hat{%
\rho}(L)$ is a vector field on $V$ one can
make use of the primitive Hopf algebra structure ($\Delta \left( L\right)
=L\otimes 1+1\otimes L$) and rewrite (\ref{leibniz}) as $L\triangleright
\left( a\cdot b\right) =\left( L_{\left( 1\right) }\triangleright a\right)
\cdot \left( L_{\left( 2\right) }\triangleright b\right) $ which provides
the module algebra condition over $\mathcal{U}_{\mathfrak{g}}$.

Realization (\ref{x1}) allows to merge the initial Lie algebra $\mathfrak{g}$ with a canonical Heisenberg algebra. Resulting algebra can be represented by the following commutation relations
\begin{equation}\label{x2}
  [L_a,L_b] = \gamma_{ab}^c L_c , \quad  [L_b, p_\nu]= (L_b)_{\nu}^\alpha p_\alpha ,\quad [p_\mu, p_\nu]=0
\end{equation}
\begin{equation}\label{x3}
   [L_a, x^\mu] =-(L_a)_{\alpha}^\mu x^\alpha , \quad [p_\nu, x^\mu]= \,1\,\delta^\mu_\nu ,\quad  [x^\mu, x^\nu]= 0
\end{equation}
The first line (\ref{x2}) represents a Lie subalgebra which can be identified as a  inhomogeneous extension $\mathfrak{ig_\rho}$  of the initial Lie algebra
$\mathfrak{g}$ with respect to the representation $\rho$. Thus the unital associative  algebra generated by the relations (\ref{x2})-(\ref{x3}) can be  introduced as a smash product  $Pol[x^1,\ldots,x^m]\rtimes U_{\mathfrak{ig_\rho}}$, where $Pol[x^1,\ldots,x^m]=U_{Ab[x^1,\ldots,x^m]}$ is the same as an enveloping algebra of the
Abelian Lie algebra $[x^\mu, x^\nu]= 0$. This former algebra can be interpreted as an extended (quantum) phase space. It contains Heisenberg algebra (quantum phase space) as a subalgebra $gen\{x^1,\ldots,x^m, p_1,\ldots,p_m\}=Pol[x^1,\ldots,x^m]\rtimes Pol[p_1,\ldots,p_m]$ which can be interpreted as an algebra of differential operators with polynomial coefficients acting  on the space $V$. Natural Hopf action of $U_{\mathfrak{ig_\rho}}$ on the module algebra $Pol[x^1,\ldots,x^m]$ is given by the commutators
(\ref{x3}): $L_a\triangleright x^\mu =-(L_a)_{\alpha}^\mu x^\alpha,
p_\nu\triangleright x^\mu= \,1\,\delta^\mu_\nu $ and provides a representation of the entire smash product algebra $Pol[x^1,\ldots,x^m]\rtimes U_{\mathfrak{ig_\rho}}$ on the space $C^\infty (V)$. This representation can be extended, in turn,  to a Hilbert space representation by introducing a suitable scalar product.

We would like to point out that while extending the Lie algebra (\ref{x2}) by adding the vector space (\ref{x3}) of some representation we exit beyond the category of Lie algebras. However, we remain in a category of associative unital algebras which include Lie algebras as a subcategory. To be more precise, obtained algebra is not an enveloping algebra of some Lie algebra. In other words the unit as a group-like element cannot belong to any Lie algebra. Consequently,  instead of bi- (Hopf) algebras one gets bi- (Hopf) algebroids. Replacing the unit by a central (primitive) Lie-algebraic element we change the structure in such a way that Lie algebra and therefore Hopf algebra are possible. But both structures are not equivalent (isomorphic) in the algebraic sense. We claim that our construction is more natural for physics since in the undeformed case it is related to Quantum Mechanics and representations (infinitesimal version) of the so-called Mackey's imprimitivity systems.

Any Drinfel'd twist in the Hopf algebra $U_{\mathfrak{ig_\rho}}$ can be used to deforme the smash product algebra $Pol[x^1,\ldots,x^m]\rtimes U_{\mathfrak{ig_\rho}}$
in two equivalent ways as described by Theorem III.1 in order to obtain new quantum phase space. In the process of twist deformation (Sec. \ref{triang}) (requiring
extension of all objects and morphisms to the category  of modules over  $\mathbb{K}[[h]]$ ring) the Hopf algebra $U_{\mathfrak{ig_\rho}}$ gets new coproduct and antipods while the underlying module algebra $Pol[x^1,\ldots,x^m]$  gets new twist deformed (noncommutative) star product:
\begin{equation}
a\star _{F}b=m\circ F^{-1}\triangleright (a\otimes b)=(\bar{F}%
_{1}\triangleright a)\cdot (\bar{F}_{2}\triangleright b)  \label{tsp}
\end{equation}%
replacing  ordinary (commutative) multiplication of scalar-valued functions. Two-cocycle condition
guarantees associativity of the corresponding twisted star-product~(\ref{tsp}).
Note that the twisted star product $\star _{F}$ is braided commutative. One
can check that using the relation: $R=F_{21}F^{-1}=F_{2}\bar{F}_{1^{\prime
}}\otimes F_{1}\bar{F}_{2^{\prime }}=R_{1}\otimes R_{2}.$ Starting from the
definition of braided commutativity (\ref{bcom}) then from definition of
star product (\ref{tsp})
\begin{eqnarray*}
\left( R_{2}\triangleright b\right) \star_F \left( R_{1}\triangleright
a\right) &=&(\bar{F}_{1}\triangleright R_{2}\triangleright b)\cdot (\bar{F}%
_{2}\triangleright R_{1}\triangleright a)= \\
(\bar{F}_{1}F_{1^{\prime }}\bar{F}_{2^{\prime \prime }}\triangleright
b)\cdot (\bar{F}_{2}F_{2^{\prime }}\bar{F}_{1^{\prime \prime
}}\triangleright a) &=&(\bar{F}_{2^{\prime \prime }}\triangleright b)\cdot (%
\bar{F}_{1^{\prime \prime }}\triangleright a)= \\
(\bar{F}_{1^{\prime \prime }}\triangleright a)\cdot (\bar{F}_{2^{\prime
\prime }}\triangleright b) &=&a\star _{F}b
\end{eqnarray*}%
the second line equality is due to the identity: $1\otimes 1=F^{-1}F=\bar{F}%
_{1}F_{1^{\prime }}\otimes \bar{F}_{2}F_{2^{\prime }}$ and the third is
using the $\cdot $ commutativity and definition of $\star _{F}$ product.

Therefore, $\mathcal{A}_{F}=\left( A,\star _{F}\right) \in _{\mathcal{U}_{%
\mathfrak{g},h}}\mathfrak{YD}^{\mathcal{U}_{\mathfrak{g},h}}$, hence $\mathcal{A}%
_{F}\rtimes \mathcal{U}_{\mathfrak{g},h}^{F}$ can be equipped with the bialgebroid structure (with all the maps as defined in \ref{thBM}). Besides the triangular deformations one can consider smash product algebras based on quasi-triangular ones. However, in this case we may not know quantum r-matrix $R$ explicitly (e.g. in the case of non-semi-simple Lie algebras) and then
we are unable to check braided commutativity of the corresponding module algebra. This happens, e.g. in the case of the celebrated $\kappa$-Poincare symmetry (see e.g. \cite{BPLuk} and references therein). Recently it has been proposed in this case to construct
Hopf algebroid which is based on Heisenberg double instead of the smash  product construction \cite{LSW}.

Many examples of the smash product algebras, for specific Lie algebras
have been already investigated before. For example, the $\kappa $%
-deformation by twists providing $\kappa $-Minkowski algebra:
\begin{equation}\label{km}
\lbrack \hat{x}^{i},\hat{x}^{j}]=0,\quad \lbrack \hat{x}^{0},\hat{x}^{i}]=%
\frac{i}{\kappa }\hat{x}^{i},
\end{equation}
as a covariant quantum space of the extended symmetries and their smash products
was investigated in \cite{sigma6} for  $\mathcal{U}_{\mathfrak{igl}\left(
n\right),h }^{F}$-inhomogeneous general linear algebra, in \cite{BP09} for $%
\mathcal{U}_{\mathfrak{ipw},h}^{F}$- Poincar\'e-Weyl algebra (one generator extension of
Poincar\'e algebra) or in \cite{MPP15} for the case of $\mathcal{U}_{\mathfrak{%
so}\left( 2,4\right),h }^{F}$ - the conformal algebra.

Such smash product algebras, called extended phase spaces, contain deformed quantum-mechanical phase space (i.e.  $\left[ P_{\mu },\hat{x}_{\nu }\right] $ commutators). The
deformation of the momenta - coordinates sector leads to the deformation of the
Poincar\'e Casimir operator and therefore to the deformation of dispersion
relations. It is due to the fact that the standard Casimir operator $P^{2}$ of the Poincar\'e algebra
does no longer satisfy:
\begin{equation}\label{xx}
\lbrack P^{2},\hat{x}_{\mu }]\neq 2P_{\mu }
\end{equation}
once the phase space is deformed. One then looks for another
invariant operator - deformed Casimir operator $C_{\kappa }$ for which~:
\begin{equation}\label{yy}
\lbrack M_{\mu \nu },C_{\kappa }]=[C_{\kappa },P_{\mu }]=0;\quad \lbrack
C_{\kappa },\hat{x}_{\mu }]=2P_{\mu }
\end{equation}%
It will lead to deformed dispersion relation of the form:
\begin{equation}
C_{\kappa }+m_{\kappa }^{2}=0
\end{equation}
(for consequences of this effect see e.g. \cite{grb}, for discussion of the triangular case, see e.g. \cite{GFiore} and references therein).

In this framework the deformation of the Poincar\'e Casimir operator (and the corresponding deformation of dispersion relations) is motivated by the use of the noncommutative coordinates assuming that  the relations (\ref{yy}) are preserved.
However from purely algebraic point of view one deals with isomorphic algebraic structures $\mathcal{A}
_{F}\rtimes \mathcal{U}_{\mathfrak{g},h}^{F}\equiv \mathcal{A}\rtimes \mathcal{U}_{\mathfrak{g}}$ therefore both commuting and noncommuting coordinates are equally justified.

We argue in this paper that distinguishing between  noncommutative coordinates can be in turn dictated  by the  choice of bialgebroid structure. In such approach  the deformed extended phase space $\mathcal{A}_{F}\rtimes \mathcal{U}_{\mathfrak{g},h}^{F}$ is not isomorphic to the undeformed one $\mathcal{A}\rtimes \mathcal{U}_{\mathfrak{g}}$ as bialgebroids, therefore the corresponding deformed coordinates are preferred.

\section{Conclusions}

The Hopf algebroids (bialgebroids) only recently have gained attention from mathematical
physics point of view. In this note we focus on one of the physically
important cases where the bialgebroid structure arises, i.e. the smash
product construction. We focus on the smash product of triangular Hopf algebra with the
Yetter-Drinfeld module algebra and then on the smash product of their twist deformed counterparts.
On the other hand we investigate the Drinfeld twist techniques
in a category of bialgebroids. We prove that these two approaches, i.e. the twisting of a
bialgebroid or constructing bialgebroid from twisted bialgebra are isomorphic in the case
of normalized cocycle twist.

As a special example we presented how to obtain Hopf algebroid from Lie
algebra. As it is known Lie algebras have special role in physics and many of them have been considered already in the Hopf algebras framework. The smash product algebras of symmetry algebra (i.e. their corresponding Hopf algebra) with the noncommutative coordinates (Hopf) module algebra are called extended phase spaces. In the case of the twist deformation, such extended
phase spaces can easily be equipped with the bialgebroid structure as shown
in Sec. E. However it is still an open issue if $\kappa $-Minkowski spacetime (\ref{km})
and $\kappa $-Poincar\'e (quasi-triangular non triangular case) smash product algebra (true $\kappa $-extended
phase space) can be equipped with the Hopf algebroid maps. Some approaches
have been made in this direction in \cite{Zagreb}, \cite{LSW}. Another two points to consider in the future would be an accommodation of the antipode map within this formalism (knowing it is possible in the case presented in Sec. D and it was already done in \cite{BM}) and an extension of the present formalism to the more general cross product construction.

Moreover, there are plenty of other examples of particular interest in physics where the
braided commutativity appears naturally, like for example commutative superalgebras (see e.g. \cite{CW}). Also the very well known Drinfeld doubles can provide examples of Hopf algebroids (see e.g. \cite{Lu}, Theorem 5.1), i.e. the smash product of the Hopf algebra and the module of its
Drinfeld double algebra is Hopf algebroid. More complete presentation of the
physically motivated Hopf algebroids as well as more interesting examples we
postpone to another paper.

\section*{Acknowledgements}

AB was supported by Polish National Science Center (NCN), project
2014/13/B/ST2/04043 and by COST (European Cooperation in Science and
Technology) Action MP1405 QSPACE. AP acknowledges the funding from the
European Union's Horizon 2020 research and innovation programme under the
Marie Sklodowska-Curie grant agreement No 660061. The authors appreciate conversations with Alessandro Ardizzoni, Isar Goyvaerts and Paolo Saracco
at the Department of Mathematics 'Giuseppe Peano' University of Turin. We are grateful to Zoran Skoda for numerous discussions and critical remarks.

\end{document}